\documentclass[10pt, reqno]{amsart}

\usepackage{a4wide}
\usepackage[english, activeacute]{babel}
\usepackage{amsmath,amsthm,amsxtra}
\usepackage{amssymb}
\usepackage{latexsym}
\usepackage{amsfonts}
\usepackage{hyperref}
\usepackage{pdftricks}

\begin{psinputs}
\usepackage{pst-plot}
\end{psinputs}

\title{Focusing mKdV breather solutions with nonvanishing boundary conditions by the Inverse Scattering Method}
\author{Miguel A. Alejo}
\address{Departament of Theoretical Biology, University of Bonn \\ Germany}
\email{miguel.alejo@uni-bonn.de}
\date{September, 2011}
\thanks{}

\chardef\bslash=`\\ 

\theoremstyle{definition}

\theoremstyle{remark}

\numberwithin{equation}{section}

\newcommand{\bel}{\begin{equation}\label}
\newcommand{\eeq}{\end{equation}}
\newcommand{\R}{\mathbb{R}}

\newcommand{\C}{\mathbb{C}}

\DeclareMathOperator{\sech}{sech}
\def\id{{\fontsize{.85em}{1.1em}\selectfont1}\normalfont\kern-.8ex1}

\newcommand{\be}{\begin{equation}}
\newcommand{\ee}{\end{equation}}
\newcommand{\ba}{\begin{equation*}}
\newcommand{\ea}{\begin{equation*}}
\newcommand{\bea}{\begin{eqnarray}}
\newcommand{\eea}{\end{eqnarray}}
\newcommand{\bee}{\begin{eqnarray*}}
\newcommand{\eee}{\end{eqnarray*}}
\newcommand{\ben}{\begin{enumerate}}
\newcommand{\een}{\end{enumerate}}

\newcommand{\eval}[2][\right]{\relax
  \ifx#1\right\relax \left.\fi#2#1\rvert}

\begin{document}

\begin{abstract}
Using the Inverse Scattering Method with a nonvanishing boundary condition, we obtain the square $k^2$  of a focusing {\it modified} Korteweg-de Vries (mKdV)  breather solution with non zero vacuum parameter $b^2$.  We are able to factorize and simplify it in order to get explicitly the associated mKdV breather solution  $k$ with non zero vacuum parameter $b$. Moreover, taking the limiting case of zero frequency, we obtain a generalization of the Double Pole solution introduced by M.Wadati {\it et al.}
\end{abstract}
\maketitle \markboth{Focusing mKdV breather solutions with nonvanishing boundary conditions by the ISM}{Miguel A. Alejo}
\renewcommand{\sectionmark}[1]{}
\section{Introduction}
The focusing {\it modified} Korteweg-de Vries  equation (mKdV for short)
\bel{geomKdV}\frac{\partial k}{\partial t} + \frac{\partial k}{\partial s^3} + 6k^2\frac{\partial k}{\partial s}=0,~~(s,t)\in\R\times\R.\eeq
appears to be relevant in a number of different physical systems (e.g. phonons in anharmonic lattices, models of traffic congestion, curve motion  and fluid mechanics between others). Indeed, it can also be considered as a canonical equation, as KdV, sine-Gordon and non linear Schr\"odinger equations.\\

Breather solutions  of mKdV \eqref{geomKdV} in the line were  found by M.Wadati \cite{W} (see also G.Lamb \cite{La}). They were rediscovered by C.Kenig, G.Ponce and L.Vega  in the proof of the discontinuity of the flowmaps associated to mKdV equation in the Sobolev spaces $H^\sigma$, constituted by functions with $\sigma$ derivatives in $L^2(\R)$ (see \cite{KePV}). Those breather solutions are defined in all the real line, vanish exponentially at infinity and, in a qualitative point of view, describe  traveling wave packets.

Indeed, they are determined by four real parameters, two of them  given by  the amplitude ($\beta$) of the envelope and the frequency ($\alpha$) of the carried wave and the other two given by  time and spatial translations (represented by $u_0,v_0$ below). It is well known that such breather solutions can be written as follows%
\bel{breather}k(s,t)= - i\frac{\partial}{\partial s}\log\left(\frac{f(u)+ig(v)}{f(u)-ig(v)}\right)=2\frac{\partial}{\partial s}\arctan\left(\frac{g(v)}{f(u)}\right),
\end{equation}
where $u=2\beta s+\gamma t+u_0$,\hspace{0.2cm} $v=2\alpha s+\delta t+v_0$,\hspace{0.2cm}$\gamma=8\beta(-\beta^2+3\alpha^2)$,\hspace{0.2cm} $\delta=8\alpha(\alpha^2-3\beta^2)$, and\footnote[1]{The $arctan$ phase in the argument of $g$ can be dropped with a suitable translation in time and space, but it will not be done  for comparison purposes.}  %
\begin{eqnarray}
\label{104}g(v)&=&\frac{\beta}{\alpha}\sin\left(v-\arctan\left(\frac{\beta}{\alpha}\right)\right), \\
\label{103}f(u)&=&\cosh(u).
\end{eqnarray} 

 This paper is devoted to the use of  the Inverse Scattering Method (ISM for short) under a nonvanishing boundary condition(NVBC shortly) devised by T.Kawata and H.Inoue \cite{IK} to obtain breather solutions of mKdV \eqref{geomKdV} which at infinity behave as a non trivial constant parameter $b$. The interest of this kind of solutions of the mKdV is related with the problem of the evolution of closed planar curves under the mKdV flow  in the following way. By the work of R.Goldstein and D.Petrich \cite{GoP} the equation \eqref{geomKdV}  is considered as the evolution equation of the curvature of a curve. By this work, closed curves can be considered as those whose associated curvature satisfies that its mean is non-zero. Moreover, those curvatures can be obtained from a solution of the focusing mKdV constructed by the (nonlinear) superposition of a constant (e.g. constant $b$) plus a traveling wave. In fact, it is posible to find some special breather solutions associated to simple closed curves that, when evolving under the mKdV flow, they create and annihilate self-intersections  (see \cite{Ale} for more information).\\
 
 Although in the literature some kind of breather solutions of the Gardner equation(also known as the extended KdV equation) have been obtained before \cite{GrSl, PeGr}, they do not contain  details on derivations of these breather solutions from the ISM scheme. Even despite the close relation between Gardner and mKdV equations (solutions of the mKdV equation with NVBC are also solutions of the Gardner equation  with zero boundary condition), breathers obtained in these references \cite{GrSl, PeGr} are far to be easily compared with the breather of M.Wadati \cite{W} they generalize. These two aspects are attained too in the present paper(for further details see section \ref{matrizDet} and \eqref{kaw112}). Finally note that the mKdV breather solutions mentioned above can be obtained alternatively using the Hirota method with a suitable selection of the wavenumbers (see the work of K.Chow and D.Lai \cite{CLai}). 
\section{Breather solutions of the focusing mKdV with nonvanishing boundary conditions}
In this section we obtain breathers of the focusing mKdV \eqref{geomKdV} with nonvanishing boundary conditions  by using the ISM  for potentials that are not trivial at infinity, introduced by T.Kawata and H.Inoue in \cite{IK}. We also recall the work of T.Au-Yeung {\it et al.}\cite{AuYF}, in which the same approach was used to obtain one and two soliton solutions with non trivial values at infinity. First, we summarize some basic results from \cite{IK} and \cite{AuYF}, necessary for our research.

\subsection{Basic results of T.Kawata and H.Inoue for the mKdV}

T.Kawata and H.Inoue considered in \cite{AKNS} a generalized AKNS eigenvalue problem  for nonvanishing potentials, which consist in the following spatial and time evolution equations\footnote{In what follows,  subindices of the form ${f}_s$ means $\partial_s f\equiv\frac{\partial f}{\partial s}$.}:

\bel{EvolX}{\bf v}_s=D(\lambda){\bf v},~~D(\lambda;s,t)=\begin{pmatrix}-i\lambda&q(s,t)\\r(s,t)&i\lambda\end{pmatrix},\eeq
\bel{EvolT}{\bf v}_t=F(\lambda){\bf v},~~F(\lambda;s,t)=\begin{pmatrix}A(\lambda)&B(\lambda)\\C(\lambda)&-A(\lambda)\end{pmatrix},\eeq
where
\bel{coefT}\begin{array}{lll}
A(\lambda) = -4i\lambda^3 - 2i\lambda qr  + rq_s  - qr_s,\\\\
B(\lambda) =  4\lambda^2q + 2i\lambda q_s + 2q^2r - q_{ss},\\\\
C(\lambda) =  4\lambda^2r - 2i\lambda r   + 2qr^2 - r_{ss}.\end{array}\eeq

Matrices $D(\lambda)$ and $F(\lambda)$ satisfy the well known integrable condition associated to equations \eqref{EvolX} and \eqref{EvolT} (i.e $\partial_{t}$\eqref{EvolX}$=\partial_s$\eqref{EvolT}):
\bel{integrabilidad}D_t - F_s + DF - FD = 0.\eeq
They seek a real potential solution $q(s, t)$ under the following boundary condition:
\bel{condInfty}q(s, t)\rightarrow b~~~ \text{as}~~~ s\rightarrow\pm\infty;~~ b^2= -\lambda_0^2,\eeq
requiring that $q(s, t)$ is sufficiently smooth and all the $s$ derivatives of $q$ tend to zero as $s\rightarrow\pm\infty$. For this purpose, they consider potentials $q(s, t)$ and $r(s, t)$ with the following nonvanishing conditions:

\bel{condPotentials}\begin{array}{lll} q(s,t) (\text{or}~~ r(s,t))\rightarrow q^{\pm} (\text{or}~~ r^{\pm})~~~ \text{as}~~~s\rightarrow\pm\infty,\\\\
q^+r^+ = q^-r^- = \lambda_0^2,\end{array}\eeq
where $q^{\pm},r^{\pm}$ and $\lambda_0^2$ are constants. Then, the spatial evolution matrix $D(\lambda;s,t)$ can be written as follows:
\begin{subequations}\label{ecuacionesDD}
\bel{ecuacionesDD1}D(\lambda;s,t)= D^{\pm}(\lambda) + \Delta D^{\pm}(s,t),\eeq\\
\bel{ecuacionesDD2}D^{\pm}(\lambda)=\lim_{s\rightarrow\pm\infty}D(\lambda;s,t)=\begin{pmatrix}-i\lambda&q^{\pm}\\r^{\pm}&i\lambda\end{pmatrix},\eeq\\
\bel{ecuacionesDD3}\Delta D^{\pm}(s,t)= D(\lambda;s,t)-D^{\pm}(\lambda)=\begin{pmatrix}0&q(s,t)-q^{\pm}\\r(s,t)-r^{\pm}&0\end{pmatrix},
\eeq
\end{subequations}
and the characteristic roots of $D^{\pm}(\lambda)$ are $\pm i\zeta$, with $\zeta=\sqrt{\lambda^2-\lambda_0^2}$. Now, they define

\bel{temp1}\begin{array}{ll}T^{\pm}(\lambda,\zeta)=\lim_{s\rightarrow\pm\infty}T(\lambda,\zeta;s),\end{array}\eeq
where \begin{equation}\nonumber T(\lambda,\zeta;s)=\begin{pmatrix}-iq_1(s)&\lambda-\zeta\\\lambda-\zeta&ir_1(s)\end{pmatrix},\end{equation}
and $q_1$, $r_1$ are suitable smooth and satisfy that $q_1(s)r_1(s)=\lambda_0^2$ for all $s\in\R$. The matrices $D^{\pm}(\lambda)$ can be diagonalized by $T^{\pm}(\lambda,\zeta)$ as

\bel{temp2}D^{\pm}(\lambda)=-i\zeta T^{\pm}(\lambda,\zeta)\sigma_3[T^{\pm}(\lambda,\zeta)]^{-1}.\eeq
Using \eqref{temp2}, they can define Jost matrices $\Phi^{\pm}$ as the solutions of \eqref{EvolX} under conditions

\bel{temp3}\Phi^{\pm}(\lambda,\zeta;s,t)\rightarrow T^{\pm}(\lambda,\zeta)J(\zeta s) ~~\text{as}~~s\rightarrow\pm\infty,\eeq
where

$$J(\zeta s)=\begin{pmatrix}e^{-i\zeta s}&0\\0&e^{i\zeta s}\end{pmatrix}.$$
Then, a scattering matrix $S$ is defined by

\bel{scat1}\Phi^{-}(\lambda,\zeta;s,t)=\Phi^{+}(\lambda,\zeta;s,t)S(\lambda,\zeta;t),\eeq
and using  relations \eqref{EvolT},\eqref{temp2}, \eqref{scat1} it is easy to derive the following condition:

\bel{temp4}S_t + SW^{-} - W^{+}S =0,~~S|_{t=0}=S_0,\eeq
where $S_0$ is given by the direct scattering, and 

\bel{Wtemp}W^{\pm}=\zeta\sigma_3\cdot\sum_{n=1}^{N}a_n\lambda^{n-1} + i(a_0/\zeta)\begin{pmatrix}-i\lambda &r^{\pm}e^{2i\zeta s}\\q^{\pm}e^{-2i\zeta s}&i\lambda\end{pmatrix}.\eeq
 The solution of \eqref{Wtemp} is easily obtained as follows

\bel{eqS}S(\lambda,\zeta;t)=e^{W^{+}t}S_0(\lambda,\zeta)e^{-W^{-}t}.\eeq
At this point, we apply this framework to the mKdV equation \eqref{geomKdV}, by  choosing  $q(s, t) = k(s, t)$ and $r(s, t) = -q(s, t)$. Then, we calculate from \eqref{Wtemp} the temporal evolution of the elements of the scattering matrix  $(S_{ij})$. For that purpose (see  \cite[p.1723]{IK}), we select $N=3$ and $a_0=a_2=0$, $a_1=2ib^2,~a_3=-4i$ in \eqref{Wtemp}, so that:

$$S_{11}(\lambda,\zeta;t)=S_{11}(\lambda,\zeta;0),\hspace*{5mm}S_{12}(\lambda,
\zeta;t)=S_{12}(\lambda,\zeta;0)e^{-4i\zeta(2\lambda^2-b^2)t},$$
$$S_{21}(\lambda,\zeta;t)=S_{21}(\lambda,\zeta;0)e^{4i\zeta(2\lambda^2-b^2)t},
\hspace*{5mm}S_{22}(\lambda,\zeta;t)=S_{22}(\lambda,\zeta;0).$$
Now, assume that zeros of the  $S_{11}(\lambda,\zeta;0)$ matrix element in the region $\Im(\zeta)>0$ are $(\lambda_j,\zeta_j),~~j=1,2$ where
\bel{raizeta} \zeta_j=\sqrt{\lambda_j^2+b^2},~~j=1,2,~~\text{for}~~~ y > s,\eeq
 and define the functions $K^{\pm}(s,y)$ as those satisfying the following three equations (for the shake simplicity, we drop the time dependence in $K^{\pm}$ and $H_c$ defined below):
\begin{subequations}\label{ecuacionesK}
\bel{ecuacionesK1}\begin{array}{ll}\frac{\partial K^{\pm}}{\partial s}(s,y)+\sigma_3\frac{\partial K^{\pm}}{\partial y}(s,y)\sigma_3+\sigma_3K^{\pm}(s,y)\sigma_3[D^{\pm}(\lambda)+i\lambda\sigma_3]
-\begin{pmatrix}0&k(s,t)\\-k(s,t)&0\end{pmatrix}K^{\pm}(s,y)=0,\end{array}\eeq
\bel{ecuacionesK2}\sigma_3K^{\pm}(s,s)\sigma_3-K^{\pm}(s,s)+\Delta D^{\pm}(\lambda;s,t)=0,\eeq
\bel{ecuacionesK3}K^{\pm}(s,y)\rightarrow0~~\text{as}~~ y\rightarrow\pm\infty,\eeq
\end{subequations}
where $D^{\pm}(\lambda)$ and $\Delta D^{\pm}(\lambda;s,t)$ are defined by \eqref{ecuacionesDD} and $\sigma_3\equiv\begin{pmatrix}1&0\\0&-1\end{pmatrix}$. Then, the Gelfand-Levitan  equation reads as follows,
\bel{GelfandEq}\begin{array}{ll}
K^+(s,y)\cdot\left(\begin{smallmatrix}0\\1\end{smallmatrix}\right) +
\sum^{2}_{j=1}\left(\begin{smallmatrix}c_j\\\tilde{c}_j\end{smallmatrix}
\right)e^{i\zeta_j(s+y)}-
H_{c}(s+y)\left(\begin{smallmatrix}0\\1\end{smallmatrix}\right)\\\\-
\int^{+\infty}_{s}K^+(s,y^{\prime})\sum^{2}_{j=1}\left(\begin{smallmatrix}
c_j\\\tilde{c}_j\end{smallmatrix}\right)e^{i\zeta_j(y+y^{\prime})}dy^{\prime}
+\int^{+\infty}_{s}K^+(s,y^{\prime})H_{c}(s+y^{\prime})\left(\begin{smallmatrix}
0\\1\end{smallmatrix}\right)dy^{\prime}=0,\end{array}\eeq
where \begin{eqnarray}     
\label{Gelfand1} c_j&=&\frac{i}{2}(\lambda_j-\zeta_j)m(\lambda_j)e^{
4i\zeta_j(2\lambda^2_j-b^2)t},\\
\label{Gelfand2} \tilde{c}_j&=&\frac{b}{2}m(\lambda_j)e^{4i\zeta_j(2\lambda^2_j-b^2)t},\\
\label{Gelfand3} m(\lambda)&=&\frac{S_{21}(\lambda,\zeta;0)}{\zeta\frac{d}{d\lambda}
S_{11}(\lambda,\zeta;0)}.
       \end{eqnarray}
We rewrite \eqref{GelfandEq} assuming:
\begin{enumerate}
        \item [a.)]Continuous component
\bel{kaw8}H_{c}(y;0)=0.\end{equation}
        \item[b.)]Zeros of $S_{11}(\lambda,\zeta)$ in the region $\Im(\zeta) >0$ are  $(\lambda_j,\zeta_j)$ and $(-\lambda_j,\zeta_j),~~j=1,2$, where
\bel{kaw87}\zeta_j=\sqrt{\lambda_j^2+b^2},\hspace*{3mm}\lambda_1=\alpha+i\beta,
\hspace*{3mm} \lambda_2=-\lambda_1^{*}=-\alpha+i\beta.\end{equation}
\end{enumerate}
Taking into account that $m(-\lambda^{*})=m(\lambda)^{*}$ (see  \cite[p.1728]{IK}), \eqref{GelfandEq} becomes

\bel{kaw88}\begin{array}{ll}
K^+(s,y)\begin{pmatrix}0\\1\end{pmatrix} + \sum^{2}_{j=1}\left(\begin{smallmatrix}-i\zeta_jm(\lambda_j)e^{4i\zeta_j(2\lambda^2_j-b^2)t}\\\\
bm(\lambda_j)e^{4i\zeta_j(2\lambda^2_j-b^2)t}\end{smallmatrix}\right)e^{i\zeta_j(s+y)}\\\\

-\int^{+\infty}_{s}K^+(s,y^{\prime})\sum^{2}_{j=1}\left(\begin{smallmatrix}-i\zeta_jm(\lambda_j)e^{4i\zeta_j(2\lambda^2_j-b^2)t}\\\\
bm(\lambda_j)e^{4i\zeta_j(2\lambda^2_j-b^2)t}\end{smallmatrix}\right)e^{i\zeta_j(y+y^{\prime})}dy^{\prime}=0.\end{array}\eeq
Choosing a representation of  $\begin{pmatrix}K^+_{12}\\K^+_{22}\end{pmatrix}$ as follows

\bel{kaw89}
\begin{pmatrix}K^+_{12}(s,y)\\K^+_{22}(s,y)\end{pmatrix}=\sum^{2}_{j=1}\begin{pmatrix}K_j(s)\\\tilde{K}_j(s)\end{pmatrix}e^{i\zeta_jy},
\eeq
then

\bel{kaw90}K^{+}(s,y)=\begin{pmatrix}K^{+}_{11}(s,y)&K^{+}_{12}(s,y)\\
K^{+}_{21}(s,y)&K^{+}_{22}(s,y)\end{pmatrix}=\begin{pmatrix}
K^{+}_{22}(s,y)&K^{+}_{12}(s,y)\\
-K^{+}_{12}(s,y)&K^{+}_{22}(s,y)
\end{pmatrix}.\end{equation}
Now, substituting equations \hypertarget{mytarget}{\eqref{kaw89}} and \hypertarget{mytarget}{\eqref{kaw90}} in \hypertarget{mytarget}{\eqref{kaw88}}, we obtain the system

\bel{kaw92}\begin{array}{ll}K_j(s)+\sum^{2}_{n=1}K_n(s)\tilde{a}_j\frac{e^{
i(\zeta_j+\zeta_n)s}}{i(\zeta_j+\zeta_n)}+\sum^{2}_{n=1}\tilde{K}_n(s)a_j\frac{e^{i(\zeta_j+\zeta_n)s}}{i(\zeta_j+\zeta_n)}+ a_je^{i\zeta_js}=0,\\\\

\tilde{K}_j(s)+\sum^{2}_{n=1}\tilde{K}_n(s)\tilde{a}_j\frac{e^{i(\zeta_j+\zeta_n)s}}{i(\zeta_j+\zeta_n)}-\sum^{2}_{n=1}K_n(s)a_j\frac{e^{i(\zeta_j+\zeta_n)s}}{i(\zeta_j+\zeta_n)}+ \tilde{a}_je^{i\zeta_js}=0,\end{array}\eeq
where
\bel{kaw93}
a_j=-i\zeta_jm(\lambda_j)e^{4i\zeta_j(2\lambda^2_j-b^2)t},\hspace*{3mm}\tilde{a}_j=bm(\lambda_j)e^{4i\zeta_j(2\lambda^2_j-b^2)t}.\eeq
which can be rewritten in a matricial form as
\bel{kaw94}
\begin{pmatrix}E&B\\-B&E\end{pmatrix}\begin{pmatrix}\vec{K}\\\vec{\tilde{K}}\end
{pmatrix}=\begin{pmatrix}\vec{A}\\\vec{\tilde{A}}\end{pmatrix},
\end{equation}
where
$\vec{K}=\begin{pmatrix}K_1\\K_2\end{pmatrix}$,\hspace*{3mm}$\vec{\tilde{K}}
=\begin{pmatrix}\tilde{K}_1\\\tilde{K}_2\end{pmatrix}$,\hspace*{3mm}$\vec{A}
=\begin{pmatrix}-a_1e^{i\zeta_1s}\\-a_2e^{i\zeta_2s}\end{pmatrix}$,\hspace*{3mm}
$\vec{\tilde{A}}=\begin{pmatrix}-\tilde{a}_1e^{i\zeta_1s}\\\-\tilde{a}_2e^{
i\zeta_2s}\end{pmatrix},$\\

\bel{kaw95}
E=(E_{jn})=\delta_{jn}+\tilde{a}_j\frac{e^{i(\zeta_j+\zeta_n)s}}{
i(\zeta_j+\zeta_n)},\hspace*{1mm}j,n=1,2,
\end{equation}
\bel{kaw96}
B=(B_{jn})=a_j\frac{e^{i(\zeta_j+\zeta_n)s}}{i(\zeta_j+\zeta_n)},\hspace*{1mm}
j,n=1,2.
\end{equation}
Defining the determinant  of the coefficient matrix by
\bel{Delta}\Delta=\begin{vmatrix}E&B\\-B&E\end{vmatrix},\end{equation}
and by using \hypertarget{mytarget}{\eqref{kaw89}}, we obtain that the solution of the system \hypertarget{mytarget}{\eqref{kaw92}} is
 
\bel{kaw98}K^{+}_{22}(s,s)=-\frac{1}{2}\frac{d}{ds}(\log\Delta).\end{equation}
From  the components of \eqref{ecuacionesK} we get, $k^2(s,t)= b^2 - 2\frac{d}{ds}K_{22}(s,s)$. Inserting \hypertarget{mytarget}{\eqref{kaw98}} into this relation, we finally obtain the expression 
\bel{kaw99}k^2(s,t)= b^2 + \frac{d^2(\log\Delta)}{ds^2}.\eeq
\subsection{The breather solution of mKdV with nonvanishing boundary value.}\label{matrizDet}
Now, our aim is to obtain an explicit expression of the focusing mKdV solution $k$ from \eqref{kaw99}. For that purpose, we rewrite the determinant \eqref{Delta} in the four dimensional case (as it corresponds to the breather case) as a product of two simpler $2\times2$ determinants. Before that, we remark the following facts about the roots ($\zeta_1,\zeta_2$) and the temporal dependence of the coefficient $S_{11}$. First, recall that in the breather case with nonvanishing boundary conditions the roots of $S_{11}(\lambda,\zeta)$ are given by
\bel{kawRoot}\zeta_j=\sqrt{\lambda_j^2+b^2},~~j=1,2,\hspace*{3mm}\lambda_1=\alpha+i\beta,\hspace*{3mm} \lambda_2=-\lambda_1^{*}=-\alpha+i\beta.\end{equation}
We express $\zeta_i,~~i=1,2$ as complex numbers as
\bel{zetas}\begin{array}{ll}\zeta_1=\tilde{\alpha}(\alpha,\beta,b)+i\tilde{\beta}(\alpha,\beta,b),\\\\
\zeta_2=-\tilde{\alpha}(\alpha,\beta,b)+i\tilde{\beta}(\alpha,\beta,b),\end{array}\eeq
where we only consider the region $\Im(\zeta)>0$ so that
\bel{kaw104}\tilde{\alpha}(\alpha,\beta,b)=\sqrt[4]{(\alpha^2-\beta^2+b^2)^2+4\alpha^2\beta^2}\cos\left(\frac{1}{2}\arctan(\frac{
2\alpha\beta}{\alpha^2-\beta^2+b^2})\right),\end{equation}
\bel{kaw105}\tilde{\beta}(\alpha,\beta,b)=\sqrt[4]{(\alpha^2-\beta^2+b^2)^2+4\alpha^2\beta^2}\sin\left(\frac{1}{2}\arctan(\frac{
2\alpha\beta}{\alpha^2-\beta^2+b^2})\right),\end{equation}
which satisfy\\
$$\tilde{\alpha}(\alpha,\beta,0)=\sqrt{\alpha^2+\beta^2}\cos\left(\arctan(\frac{\beta}{\alpha})\right)=\Re(\alpha+i\beta)=\alpha,$$
$$\tilde{\beta}(\alpha,\beta,0)=\sqrt{\alpha^2+\beta^2}\sin\left(\arctan(\frac{\beta}{\alpha})\right)=\Im(\alpha+i\beta)=\beta.$$
Now, by using \eqref{zetas}, the exponents of the temporal dependence given by \eqref{kaw93} are rewritten as

\bel{depTempo}\begin{array}{ll}4i\zeta_1(2\lambda_1^2-b^2)=-\tilde{\gamma}(\alpha,\beta,b)+i\tilde{\delta}(\alpha,\beta,b),\\\\

4i\zeta_2(2\lambda_2^2-b^2)=-\tilde{\gamma}(\alpha,\beta,b)-i\tilde{\delta}(\alpha,\beta,b),\end{array}\eeq
and, again, we only consider the region $\Im(\zeta)>0$ so that
\bel{kaw106}\tilde{\gamma}(\alpha,\beta,b)=8\tilde{\beta}(3\tilde{\alpha}^2-\tilde{\beta}^2)-12b^2\tilde{\beta},\eeq
\bel{kaw107}\tilde{\delta}(\alpha,\beta,b)=8\tilde{\alpha}(\tilde{\alpha}^2-3\tilde{\beta}^2)-12b^2\tilde{\alpha},\eeq
which satisfy 
\bel{velGama}\tilde{\gamma}(\alpha,\beta,0)=8\beta(3\alpha^2-\beta^2)=\gamma,\eeq
\bel{velDelta}\tilde{\delta}(\alpha,\beta,0)=8\alpha(\alpha^2-3\beta^2)=\delta.\eeq
For the sake of simplicity and without loss of generality, we rename $\tilde{\alpha}, \tilde{\beta},  \tilde{\gamma}, \tilde{\delta}$ as $\alpha,\beta,\gamma, \delta$, respectively. Now, we are able to rewrite the determinant $\Delta$ as  a product of two simpler $2\times2$ determinants. 
\bel{kaw110}\begin{array}{lll}
\Delta &=&\det\left((I-iM-ibN)\cdot(I+iM-ibN)\right)\\\\
&=&\det(I-iM-ibN)\det(I+iM-ibN)\\\\
&=&\Re \left\{det(I-iM(s,t)-ibN(s,t))\right\}^2+\Im\left\{det(I-iM(s,t)-ibN(s,t))\right\}^2,\end{array}
\eeq
where 

\bel{kaw111}\begin{array}{ll}I\equiv\id_{2x2}=\begin{pmatrix}
        1&0\\
        0&1\end{pmatrix}, 
        \hspace*{3mm}   
        M(s,t)=\begin{pmatrix}
        \frac{-me^{2i(\alpha+i\beta)s+(-\gamma+i\delta)t}}{2}&\frac{i(\alpha+i\beta)me^{-2\beta
s+(-\gamma+i\delta)t}}{2\beta}\\
        \frac{i(-\alpha+i\beta)m^{*}e^{-2\beta
s-(\gamma+i\delta)t}}{2\beta}&
\frac{-m^{*}e^{2i(-\alpha+i\beta)s-(\gamma+i\delta)t}}{2}
\end{pmatrix},\\\\
N(s,t)=\begin{pmatrix}
        \frac{me^{2i(\alpha+i\beta)s+(-\gamma+i\delta)t}}{2(\alpha+i\beta)}&\frac{-ime^{-2\beta
s+(-\gamma+i\delta)t}}{2\beta}\\
        \frac{-im^{*}e^{-2\beta
s-(\gamma+i\delta)t}}{2\beta}&\frac{m^{*}e^{2i(-
\alpha+i\beta)s-(\gamma+i\delta)t}}{2(-\alpha
+i\beta)}\end{pmatrix}.
\end{array}\eeq
Then, substituting \eqref{kaw110} in \eqref{kaw99}, and resorting to the identity
\bel{identlogarct}2\frac{\partial}{\partial s}\arctan(\frac{z}{w})=i\frac{\partial}{\partial s}\log(\frac{w-iz}{w+iz}),\eeq 
we get
\bel{kaw111a}\begin{array}{ll} k^2(s,t)=b^2
  + \frac{d^2}{ds^2}\log\left(\Re \left\{det(I-iM(s,t)-ibN(s,t))\right\}^2+\Im\left\{det(I-iM(s,t)-ibN(s,t))\right\}^2\right)\\\\
 =\left(b +i\frac{\partial}{\partial s}\log\left[\frac{\Re\left\{ det(I-iM(s,t)-ibN(s,t))\right\}-i\Im \left\{det(I-iM(s,t)-ibN(s,t))\right\}}{\Re\left\{det(I-iM(s,t)-ibN(s,t))\right\}+i\Im\left\{det(I-iM(s,t)-ibN(s,t))\right\}}\right]\right)^2\\\\
  =\left(b +2\frac{\partial}{\partial x}\arctan\left[\frac{\Im\left\{det(I-iM(s,t)-ibN(s,t))\right\}}{\Re\left\{ det(I-iM(s,t)-ibN(s,t))\right\}}\right]\right)^2,
\end{array}\end{equation}
which gives  directly the matricial expression for the breather solution of the focusing mKdV with nonvanishing boundary value:\\\\
\bel{kaw112}
 k(s,t)=b +2\frac{\partial}{\partial s}\arctan\left[\frac{\Im\left\{det(I-iM(s,t)-ibN(s,t))\right\}}
 {\Re\left\{det(I-iM(s,t)-ibN(s,t))\right\}}\right].\eeq\\
In fact, it is possible to calculate from \eqref{kaw112} the explicit expression for $k$. For that, we first calculate the determinant in \eqref{kaw112}(we write $m=|m|e^{i\phi}$):
\bel{detexplicito}\begin{array}{lll}f(s,t)+ig(s,t)\equiv\det(I-iM(s,t)-ibN(s,t))\\\\
=1 +
(1-\frac{b^2}{\alpha^2+\beta^2})\frac{\alpha^2|m|^2}{
4\beta^2}e^{-4\beta
s-2\gamma t}+i\frac{m}{2}e^{2i(\alpha+i\beta)s+(-\gamma+i\delta)t}+i\frac{m^{*}}{2}e^{2i(-\alpha+i\beta
)s-(\gamma+i\delta)t}\\\\
+i\frac{b\alpha^2|m|^2}{2\beta(\alpha^2+\beta
^2))}e^{-4\beta s-2\gamma t}+\frac{b|m|}{ \alpha^2+\beta^2}e^{-2\beta
s-\gamma t}(\alpha\sin(2\alpha s+\delta t+\phi)-\beta\cos(2\alpha s+\delta t+\phi)),\end{array}
\eeq
so that

\begin{subequations}\label{pImag}
\begin{align}g(s,t)
=&\frac{m}{2}e^{2i(\alpha+i\beta)s+(-\gamma+i\delta)t}+\frac{m^{*}}{2}e^{2i(-\alpha+i\beta)s-(\gamma
+i\delta)t}+\frac{b\alpha^2|m|^2}{2\beta(\alpha
^2+\beta^2))}e^{-4\beta s-2\gamma t},\\
\nonumber f(s,t)
=&1+(1-\frac{b^2}{\alpha^2+\beta^2})\frac{\alpha^2|m|^2}{
4\beta^2}e^{-4\beta s-2\gamma t}
\\\label{pReal}&+\frac{b|m|}{\alpha^2+\beta^2}e^{-2\beta
s-\gamma t}(\alpha\sin(2\alpha x+\delta
t+\phi)-\beta\cos(2\alpha s+\delta t+\phi)).\end{align}\end{subequations}
By defining
$e^{-2\psi}=(1-\frac{b^2}{\alpha^2+\beta^2})\frac{\alpha
^2|m|^2}{4\beta^2}$, the expression of $g~\text{and}~ f$ simplifies to
\begin{subequations}\label{RePimag0}
\begin{align}\label{RePimag01} g(s,t)=&2e^{-2\beta s-\gamma t-\psi}\tilde{g}(s,t),\\
\label{RePimag02}f(s,t)
=&2e^{-2\beta s-\gamma t-\psi}\tilde{f}(s,t),
\end{align}
\end{subequations}
where
\begin{subequations}\label{RePimag1}
\begin{align}\nonumber \tilde{g}(s,t)&=\frac{\beta}{\alpha}\sqrt{\frac{\alpha^2+
\beta^2}{\alpha^2+\beta^2-b^2}}\cos(2\alpha s+\delta t+\phi) \\
\label{RePimag11}& +\frac{b\beta}{\alpha^2+\beta^2-b^2}\left(\cosh\left(2\beta s+\gamma t+\psi\right)-\sinh\left(2\beta s+\gamma t+\psi\right)\right),
\\
\label{RePimag21}\tilde{f}(s,t)&=\cosh(2\beta s+\gamma t+\psi)+\frac{b\beta}{\alpha\sqrt{\alpha^2+\beta^2-b^2}}\sin(2\alpha s+\delta t+\phi-\arctan(\frac{\beta}{\alpha})).
\end{align}
\end{subequations}
Hence, the explicit expression for the breather solution of the focusing mKdV with nonvanishing boundary value $b$ (or {\it b-breather}) is:\\
\bel{kaw115}k(s,t)= b  + 2\frac{\partial}{\partial
x}\arctan\left[\frac{\Im\left[f(s,t)+ig(s,t)\right]}{\Re\left[f(s,t)+ig(s,t)\right]}\right]= b  + 2\frac{\partial}{\partial
x}\arctan\left[\frac{\tilde{g}(s,t)}{\tilde{f}(s,t)}\right],\eeq
\noindent with $\tilde{f},\tilde{g}$ given by \eqref{RePimag1}. In the formal limit $b\rightarrow0$, \eqref{kaw115} is reduced to the well known breather solution \eqref{breather} of the focusing mKdV equation (see \cite{W}, up to translations in time and space). Even more, it is possible to obtain the generalization of the Double Pole solution presented by M.Wadati {\it et al.} \cite{OW} with  a nonvanishing boundary value at infinity. For that, do the translation $\tilde{v}=v+\arctan(\frac{\beta}{\alpha})$, where $v=2\alpha s+\delta t+\phi$ is the  argument of the oscillatory functions in \eqref{RePimag1}, and  calculate the formal limit $\alpha\rightarrow0$. Such generalization is given by the explicit formula 
\bel{SolpoloDoble}k(s,t)= b + 2\frac{\partial}{\partial s}\arctan\left(\frac{\tilde{G}(s,t)}{\tilde{F}(s,t)}\right)
,\eeq
where
\begin{align}
\nonumber\tilde{G}(s,t)&=\frac{\beta(1-2\beta(s-6(2\beta^2+b^2)t))}{\sqrt{\beta^2-b^2}}+\frac{b\beta\left(\cosh(u)-\sinh(u)\right)}{\beta^2-b^2},
\\
\nonumber\tilde{F}(s,t)&=\cosh(u)+\frac{2b\beta(s-6(2\beta^2+b^2)t))}{\sqrt{\beta^2-b^2}},\hspace{0.7cm}u=\beta(2s-4(2\beta^2+3b^2)t).\end{align}
 Taking into account the point-wise convergence of \eqref{SolpoloDoble} to $b$ when time goes to\footnote{The case $t\rightarrow-\infty$ is equivalent.} $t\rightarrow+\infty$, it is possible to guess its asymptotic form at the mentioned limit, which is %
\bel{asintbDP} k(s,t)\approx b + \frac{2\beta^2}{\sqrt{\beta^2+b^2}} \sech (u-\delta_{+}) - \frac{2\beta^2}{\sqrt{\beta^2+b^2}}\sech(u-\delta_{-}), \eeq
with 
\bel{fasebDP}\delta_{\pm}=\beta\log\left(12\beta(\frac{\beta^2-b^2}{\beta^2+b^2})^{\pm\frac{1}{2}}(4\beta^2+2b^2)t\right).\eeq
The phase $\delta_{\pm}$  determines how evolves  the distance between the soliton and the antisoliton of   \eqref{SolpoloDoble}.
\phantom{a}\\
\begin{figure}[!htb]
\centering
\includegraphics[width=13.0cm,height=5cm]{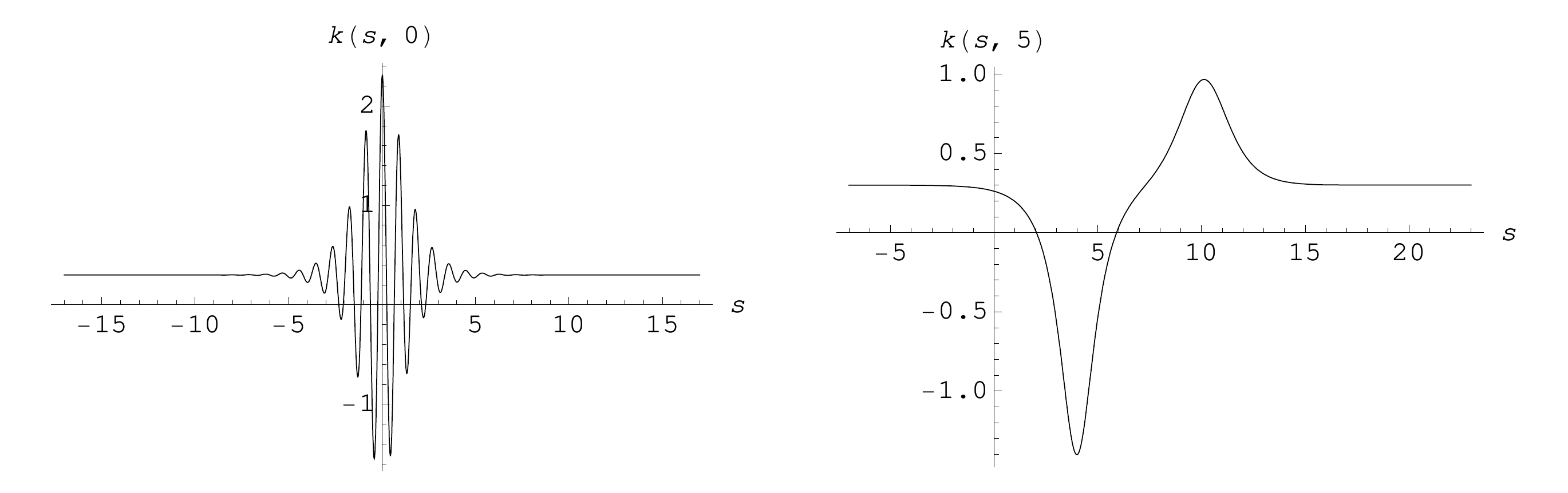}
\small{\caption{Left: Breather solution \eqref{kaw115} with $\alpha=7,\beta=1,b=0.3$ at $t=0$.~
Right: Double Pole solution \eqref{SolpoloDoble}  with $\beta=1,~b=0.3$ at $t=5$.\label{BreatherPolo}}}
\end{figure}
\phantom{a}\\
\bigskip
\section{Summary and remarks}
In this paper we have obtained  the breather solution of the focusing mKdV equation with nonvanishing boundary conditions at infinity \eqref{kaw115} by using the inverse scattering method for  potentials that are not trivial at infinity as it was devised by T.Kawata and H.Inoue in \cite{IK}.  As far as we know, it has not been reported before a systematic work on the obtention of this kind of  breather solutions of mKdV under the ISM. These solutions play an important role in the construction of closed curves with localized perturbations,  which evolve  under the mKdV flow of curves (see \cite{ Ale}).  We have also generalized the Double Pole solution found by M.Wadati {\it et al.} \cite{OW} to the case when it takes non trivial values $b$ at infinity. We have shown that even in this generalization, the distance between  humps grows proportionally to $\log(t)$, as the formula \eqref{fasebDP} shows. Moreover, the associated closed curve to this (Double Pole curvature) solution is a closed curve with two loops. These two loops enclose asymptotically the same area, they point in- and outward respectively the closed curve, travel in the same direction and the distance between them grows slowly (proportionally to $\log(t)$ as $t$ goes to $\infty$) (see  \cite{Ale}). We  think that  the asymptotic property \eqref{fasebDP} could be useful to check the accuracy of   numerical methods (e.g. difference and pseudo spectral methods) for $t$ big enough, as it was  shown  in \cite{AleGV}  when $b=0$.

Finally we would like to remark that it is possible to obtain new solutions of the defocusing mKdV equation
\bel{defmKdV}\frac{\partial k}{\partial t} + \frac{\partial k}{\partial s^3} - 6k^2\frac{\partial k}{\partial s}=0,~~(s,t)\in\R\times\R.\eeq
 from the focusing mKdV breather solutions in \eqref{kaw115} with a special choice of their free parameters $b,\beta,\alpha$. With this goal in mind, we make the following changes of parameters in \eqref{kaw115}: $b~ \text{becomes}~ ib,~~\beta~ \text{becomes}~-\beta$ and $\alpha~ \text{becomes}~i\alpha$. Then, we get a new purely complex  solution $i\tilde{k}(s,t)$ of the focusing mKdV equation \eqref{geomKdV}. Hence $\tilde{k}(s,t)$ is a real and regular two soliton solution of the defocusing mKdV equation \eqref{defmKdV} with nonvanishing boundary value $b$ at infinity. With other selections of the parameters, $b, \beta, \alpha$ and with the same procedure, we obtain different complex and regular or singular (depending on the selected parameters) solutions of the equation \eqref{defmKdV} with nonvanishing boundary value $b$ at infinity.

\end{document}